\documentclass[showpacs,pre,preprint,onecolumn]{revtex4}%
\usepackage{amsfonts}
\usepackage{amsmath}
\usepackage{amssymb}
\usepackage{graphicx}%
\begin{document}
\title{Alternans amplification following a two-stimulations protocol in a
one-dimensional cardiac ionic model of reentry: from annihilation to
double-wave quasiperiodic reentry.}
\author{P. Comtois}
\affiliation{Institute of Biomedical Engineering, Universit\'{e} de Montr\'{e}al and
Research Centre, H\^{o}pital du Sacr\'{e}-Coeur, 5400 Gouin W., Montr\'{e}al,
Qu\'{e}bec, Canada, H4J 1C5.}
\author{A. Vinet}
\affiliation{Physiology Dept. and Institute of Biomedical Engineering, Universit\'{e} de
Montr\'{e}al and Research Centre, H\^{o}pital du Sacr\'{e}-Coeur, 5400 Gouin
W., Montr\'{e}al, Qu\'{e}bec, Canada, H4J 1C5.}

\begin{abstract}
Electrical pacing is a common procedure that is used in both experimental and
clinical settings for studying and/or annihilating anatomical reentry. In a
recent study [Comtois and Vinet, Chaos \textbf{12}, 903 (2002)], new ways to
terminate the one-dimensional reentry using a simple protocol consisting of
only two stimulations were discovered. The probability of annihilating the
reentrant activity is much more probable by these new scenarios than by the
usual local unidirectional block. This paper is an extension of the previous
study in which the sensitivity of the new scenarios of annihilation to the
pathway length is studied. It follows that reentry can be stopped over a
limited interval of the pathway length and that increasing the length beyond
the upper limit of this interval yields to a transition to sustained
double-wave reentry. A similar dynamical mechanism, labeled alternans
amplification, is found to be responsible for both behaviors.

\end{abstract}
\pacs{87.19.Hh, 05.45.-a}
\keywords{Cardiac reentry, stimulation procotol, multistability}
\maketitle

\section{Introduction}

The picture of a fixed waveform traveling at constant speed around a ring of
excitable tissue, is still a common representation of functional reentry in
the clinical setting, particularly in reference to common atrial flutter
\cite{Mines1914_On,Frame1988_Os,Frame1991_Sp,Pinto1993_En,Jalil2003_Ex}.
However, the findings that complex reentries are possible even in\ a simple
homogeneous one-dimensional ionic loop model and that their occurrence is
dependent on the steepness of the restitution\ curve of the action potential
duration has altered the current understanding of the phenomena
\cite{Courtem1993_In,Vinet1994_Th,Vinet1999_Me,Vinet2000_Qu}, in which reentry
was postulated to remain stable and periodic as long as there was a minimal
excitable gap ahead of the wavefront. These findings have also altered the
thinking about the effect of antiarrhythmic drugs.
\cite{Garfink2000_Pr,Qu2000_Me}.

Overdrive pacing using a transvenously inserted catheter in the right atrium
is a standard clinical procedure to interrupt atrial flutter. \ It is very
successful, particularly when it is applied in conjunction with the
administration of class I or III antiarrhythmic drug
\cite{DellaBe1991_Us,Heldal1993_Ef} The use of rapid pacing is likely to
increase with the implant of permanent single or dual site stimulator for the
prevention of atrial tachycardias\cite{Prakash1997_Ac}. However, the mechanism
by which overdrive pacing interrupts reentry and the electrophysiological
parameters of the reentry circuit that may determined an optimal choice of
parameters for the pacing protocol are not understood.

As a first step to improve the pacing algorithm, we have previously studied of
a simple protocol of stimulation consisting of two electrical stimuli applied
in the pathway of a periodic reentry \cite{Comtois2002_Re}. New scenarios of
reentry annihilation were identified, different from the classical
unidirectional block \cite{Quan1991_Te,Shaw1995_Th,Fei1996_As}, which is still
considered to be the most important mode of termination.

These alternative scenarios of reentry annihilation follow from a
spatiotemporal process that we have called alternans amplification
\cite{Comtois2002_Re}. The first objective of this paper is to understand the
effect of length of the reentry pathway on these scenarios of annihilation. We
also show that beyond a critical length of the reentry pathway, alternans
amplification induces a transition double-wave reentry instead of annihilation.

\section{Models and methods}

Results obtained with two different models are presented. The first model
(ionic loop: IL) is\ a one-dimensional reaction diffusion system, using a
cardiac ionic model to represent the transmembrane currents. The second model
is an integral-delay equation (ID) based on the local properties of
propagation and repolarization.

\subsection{Ionic loop model}

The well-known monodomain cable equation for an 1D homogeneous excitable
cardiac tissue embedded in an unbounded external medium of negligible
resistivity is:\newline%
\begin{equation}
\frac{1}{\rho}\frac{\partial^{2}V}{\partial x^{2}}=S\left(  C_{M}%
\frac{\partial V}{\partial t}+I_{ion}+I_{stim}\right)  , \label{cable}%
\end{equation}
\newline where $V$ is the transmembrane potential (mV), $C_{m}$ is the
membrane capacitance ($1\ \mu$Fcm$^{-2}$), $S$ the surface-to-volume ratio
($0.4\ \mathcal{\mu}$m$^{-1}$, assuming cylindrical cells with radius of
$5\ \mathcal{\mu}$m) and $\rho$ is the intracellular resistivity
($200\ \Omega$cm). $I_{ion}$ is given by a modified version of the
Beeler-Reuter model (MBR) of the cardiac cell membrane, whose details and
space-clamp dynamics are given in \cite{Vinet1994_Ex}.

For each time step ($\Delta t=2\mu s)$, the system becomes a second order
ordinary differential equation that is computed with a Galerkin finite element
method projected on a linear basis function and a regular spatial mesh
($\Delta x=50\
\operatorname{\mu m}%
$)\cite{Vinet1994_Th}. The resulting tridiagonal linear system of equations is
solved with a simplified LU decomposition method. The choice of $\Delta t$ and
$\Delta x$ is motivated by the fact that depolarization is the stiffest part
of the process. Programs were written in Fortran77 and ran on SGI workstations
(Silicon Graphics).

Following the generation of an action potential, two quantities are measured
at each site to analyze the propagation: the activation time ($T_{act}(x)$)
and repolarization time ($T_{rep}(x)$). $T_{act}(x)$ corresponds to the onset
of the action potential and is defined as the moment at which $dV/dt$ reaches
its maximum during the upstroke of the action potential. $T_{rep}(x)$ is meant
to indicate the time from which a new action potential can be generated by an
incoming activation front or an external stimulus. A large set of simulations
of sustained reentry in loops of different lengths has showed that an active
propagating response was generated if stimulation was applied at least$\ 30$
ms\ after the$\ -50$ mV post-upstroke downcrossing in repolarization.
Accordingly, this instant (i.e. 30 ms after $-50$ mV crossing) is taken as
$T_{rep}(x)$. The action potential ($A\left(  x\right)  $) is defined as the
time interval from $T_{act}(x)$ to $T_{rep}(x)$, such that $T_{rep}%
(x)=T_{act}(x)+A(x)$. The diastolic interval ($D\left(  x\right)  $)
associated to an activation occurring at $T_{act}(x)$ is defined as the time
between the previous $T_{rep}(x)$ and current $T_{act}(x)$. With these
definitions, a site is excitable if $D\left(  x\right)  \geq0$.

\subsection{Integral-delay model}

The integral-delay model used in this study is an extension of a previous
formulation that was developed to describe sustained unidirectional
propagation on the loop
\cite{Courtem1996_A,Vinet2000_Qu,Comtois2002_Re,comtoispre2003}. A first
relation $A^{SP}(D)$ gives the duration of the action potential as a function
of $D$ in the space-clamped configuration. If the nodes were disconnected from
their neighbors, the repolarization time following an activation occurring at
$T_{act}(x)$ would be:%
\[
T_{rep}^{SP}\left(  x\right)  =T_{act}\left(  x\right)  +A^{SP}\left(
D\left(  x\right)  \right)
\]
provided that $D\geq0$. If $D<0$, the node is unexcitable. The actual
repolarization time of a node at position $x$ is expressed as a weighted
average of $T_{rep}^{SP}\left(  x\right)  $ over a symmetric neighborhood of
length $2\alpha$, i.e.:%
\[
T_{rep}\left(  x\right)  =\int_{-\alpha}^{\alpha}w\left(  u\right)
T_{rep}^{SP}\left(  x+u\right)  du
\]
where , $w\left(  x\right)  =\vartheta^{-1}\exp\left(  -\beta x^{2}\right)  $
with ($\vartheta=0.5$, $\beta=800$ cm$^{-2}$, $\alpha=0.5$ cm), is the
weighting function representing the effect of resistive coupling on the
repolarization phase. $\beta$ controls the spatial decay of the weighting
function and $\vartheta$ is a normalization coefficient such that
$\int_{-\alpha}^{\alpha}w\left(  u\right)  du=1$. The calculation of $T_{rep}$
associated to one excitation is performed at each node at the next instance
when it is stimulated by an incoming front or an external stimulus. At this
moment, the $T_{rep}^{SP}$ associated to the previous excitation of each point
of the neighborhood\ are collected and averaged to produce $T_{rep}$. In this
way, a front whose propagation stops at some location still produces a
continuous distribution of $T_{rep}$ \ around the region of block since
$T_{rep}$ is an weighted average of the $T_{rep}^{SP}$ \ of the sites excited
by the front that is blocked and of those that were not reached by that front
and still have the $T_{rep}^{SP}\left(  x\right)  $ associated to their
previous excitation.\ It provides at once a representation of the acceleration
of repolarization of the excited cells induced by the load of those that are
not excited, and of the prolongation of repolarization in those that were not
excited by the electrotonic depolarization induces by the proximal excited cells.

Once the $T_{rep}$ associated to the last excitation that we label
$T_{rep}^{last}(x),$ has been computed, the diastolic interval associated to
the current stimulation, which take place at the time $T_{act}(x)$, is
calculated as:%

\begin{equation}
D\left(  x\right)  =T_{act}\left(  x\right)  -T_{rep}^{last}(x)\label{mnIDE}%
\end{equation}
If $D\geq0,$the stimulus produces an action potential$,$ which propagates on
both side with a conduction time $c_{t}(D(x)),$ and reached the neighbouring
nodes at the time $T_{act}\left(  x\pm\Delta x\right)  =T_{act}\left(
x\right)  +\Delta x$ $c_{t}(D(x))$. If $D<0$, the point is not activated, and
its $T_{rep}^{SP}\left(  x\right)  $ is not changed.

The\ integral-delay model was originally developed to represent the
propagation of \ a single activation front during reentry, without external
stimulation. In this context, $T_{rep}^{last}(x)$ is always the repolarisation
associated to the previous passage of the activation front \ and it can be
written as $T_{rep}(x-L)=\int_{-\alpha}^{\alpha}w\left(  u\right)
T_{rep}^{SP}\left(  x-L+u\right)  du=\int_{-\alpha}^{\alpha}w\left(  u\right)
(T_{act}\left(  x-L+u\right)  +A^{SP}\left(  D\left(  x-L+u\right)  \right)
du$. Similarly, $T_{act}\left(  x\right)  =T_{act}\left(  x-L\right)
+\int_{o}^{L}c_{t}(D(u))du$. With these relations, eq. \ref{mnIDE}, becomes:%
\begin{align}
D\left(  x\right)   &  =T_{act}\left(  x\right)  -T_{rep}\left(  x-L\right)
\label{mIDE}\\
&  =T_{act}\left(  x-L\right)  -\int_{-\alpha}^{\alpha}w\left(  u\right)
T_{act}\left(  x-L+u\right)  du\nonumber\\
&  +\int_{0}^{L}c_{t}(D(u))du-\int_{-\alpha}^{\alpha}w\left(  u\right)
A^{SP}\left(  D\left(  x-L+u\right)  \right)  du\nonumber
\end{align}
If the conduction time is constant $\int_{-\alpha}^{\alpha}w\left(  u\right)
T_{act}\left(  x-L+u\right)  du=T_{act}\left(  x-L\right)  ,$ and eq.
\ref{mnIDE} is equivalent to the version of the ID model introduced in
\cite{Vinet2000_Qu}. In fact this version neglects the effect of the delay of
propagation in calculating the effect of coupling on repolarization. If
\ $w\left(  u\right)  $ is taken as a $\delta$ function, which is equivalent
to ignoring the effect of coupling, eq. \ref{mnIDE} \ corresponds to
the\ version of \ the integral-delay model of Courtemanche and al.\newline The
simulation of the ID model were performed using%

\begin{equation}
c_{t}\left(  D\right)  =c_{t,\min}+17.645e^{-D/2.79}+12.40e^{-D/15.34}
\label{eq_dispersion}%
\end{equation}
where $c_{t,\min}=14.09$ ms/cm, $D$ is in ms and $c_{t}$ is in ms/cm, and
\begin{equation}
A^{SP}\left(  D\right)  =42.23+\left[  A_{0}-B_{0}e^{-D/\tau}\right]
\frac{\left(  D+30\right)  ^{4.92}}{F^{4.92}+\left(  D+30\right)  ^{4.92}}
\label{eq_restitution}%
\end{equation}
with $A_{0}=225.22$ ms, $B_{0}=105.74$ ms, $\tau=94.71$ ms, and $F=59.38$ ms
\cite{Vinet2000_Qu,Comtois2002_Re}. These functions were obtained by fitting
the data gathered from different regimes of propagation (free running periodic
and QP reentry obtained with the IL model)\cite{Vinet2000_Qu}. Computations
were performed with a spatial discretization of $\Delta x=50$ $%
\operatorname{\mu m}%
$, as in the ionic model.

\subsection{Protocol of stimulation\label{protocol}}

For the IL model, $I_{stim}$ consists of a 2.5 ms current pulse applied over
an interval of 450 $%
\operatorname{\mu m}%
$ with amplitude of 60 $\mathrm{\mu}$A/cm$^{2}$. Dual stimulations were
applied on periodic reentry. The timing of \ stimuli was controlled by two
parameters: $\Delta_{F-S1}$, the time interval between the last activation at
the center of the stimulated area and the onset of the first stimulus $S_{1}$;
$\Delta_{S1-S2}$ , the time interval between the onset of $S_{1}$ and that of
the second stimulus $S_{2}$

For\ the ID\ model, the value of $D$, defines the response of the nodes that
are stimulated. If $D<0$, the stimulus is applied in the refractory period and
doest not have an effect. If $D\geq0$, the stimulus depolarizes the tissue,
defining $A^{SP}(D)$, and induces bi-directional propagation. As for the IL
model the stimulation covers 450 $%
\operatorname{\mu m}%
$.

\section{Results}

The post-stimuli dynamics are constrained by the steady states of the system.
For loops longer than a minimum length $L_{\min}$, sustained reentries are
stable attractor of the system. These sustained reentries can be either
periodic (period-1) or quasiperiodic (QP), and hold a single (SW), two (DW) or
more traveling activation fronts. Table \ref{table_attract} lists the stable
solutions of both the ID and IL models for $L<3L_{\min}$. The number and
nature of the sustained reentries change with $L$ and condition the outcomes
of the stimulations.%

\begin{table}[tbp] \centering
\begin{tabular}
[c]{|l|ll|}\hline
Interval & Reentry type & \\\hline\cline{1-2}%
$L$ \ (cm) & SW (single wave) & \multicolumn{1}{|l|}{DW(double wave)}%
\\\hline\hline
$0\leq L<L_{\min}=12.8$ & $-$ & \multicolumn{1}{|l|}{$-$}\\\hline
$L_{\min}\leq L<L_{\min}^{\text{mode-1}}=15.2$ & QP, mode-0 &
\multicolumn{1}{|l|}{$-$}\\\hline
$L_{\min}^{\text{mode-1}}\leq L<L_{crit}=19.5$ & $%
\begin{array}
[c]{c}%
\text{QP},\text{mode-0}\\
\text{QP},\text{mode-1}%
\end{array}
$ & \multicolumn{1}{|l|}{$-$}\\\hline
$L_{crit}\leq L<2L_{\min}=25.6$ & period-1 & \multicolumn{1}{|l|}{$-$}\\\hline
$2L_{\min}\leq L<2L_{\min}^{\text{mode-1}}=30.4$ & $\text{period-1}$ &
\multicolumn{1}{|l|}{$\text{QP},\text{mode-0}$}\\\hline
$2L_{\min}^{\text{mode-1}}\leq L<2L_{crit}=39$ & period-1 &
\multicolumn{1}{|l|}{$%
\begin{array}
[c]{c}%
\text{QP},\text{mode-0}\\
\text{QP},\text{mode-1}%
\end{array}
$}\\\hline
$2L_{crit}\leq L<3L_{\min}$ & $\text{period-1}$ &
\multicolumn{1}{|l|}{$\text{period-1}$}\\\hline
\end{tabular}
\caption{Stable reentries of the IL  model as a function of the length  of the loop.\label{table_attract}}%
\end{table}%

\begin{figure}
[ptb]
\begin{center}
\includegraphics[
height=3.76in,
width=3.2128in
]%
{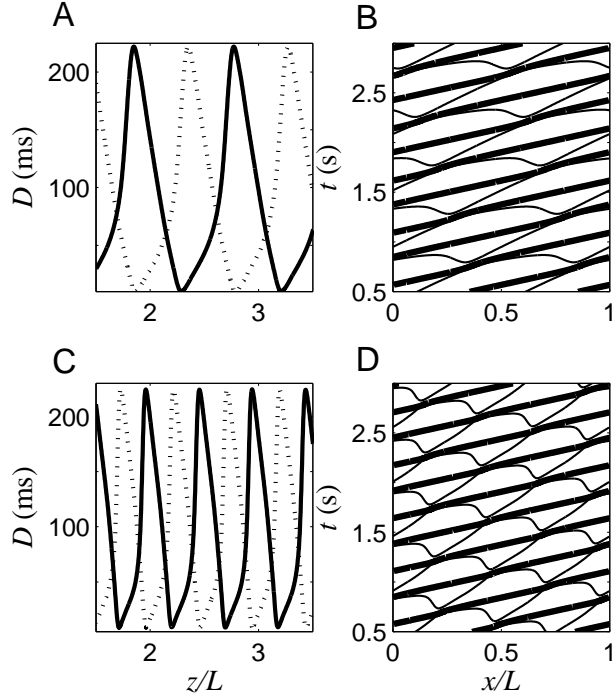}%
\caption{IL model: the mode-0 (panels A-B) and mode-1 DW \ reentry (panels
C-D) coexisting at $L=34$ cm. In this case, two fronts $F_{1}$ and $F_{2}$
travels along the loop. \textbf{Left panels:} the ordinate gives the value of
$D$ at which activation occurs at each site when it is reached by the
activation fronts ($F_{1}$, continuous line and $F_{2}$, dotted line). Each
site is activated first by $F_{1}$, and later by $F_{2\text{.}}$The abscissa
is the distance travelled by the fronts from the same arbitrary reference
point (divided by $L$). \textbf{Right panels:} successive $T_{act}\left(
x\right)  $ (thick line) and $T_{rep}\left(  x\right)  $ (thin line) at each
site. Each successive set of lines corresponds alternatively to the
$T_{act}\left(  x\right)  $ and $T_{rep}\left(  x\right)  $ profiles
associated to $F_{1}$and $F_{2}$. The locations of $F_{1}$ and $F_{2\text{ }}%
$at any time $t$ are the two positions where $T_{act}\left(  x\right)  =t$.
The successive values of \ $D$ at one location are the differences between
$T_{act}\left(  x\right)  $ and the preceding $T_{rep}\left(  x\right)  $.}%
\label{quasi_sust}%
\end{center}
\end{figure}

QP reentries are characterized by a spatial oscillation of $D$, with a
wavelength $\Lambda$ that is an irrational fraction of $L$ (fig.
\ref{quasi_sust}A and C). For both SW and DW reentries, there is an interval
of $L$ in which two different QP solutions coexist. These solutions, labeled
mode-0 and mode-1, have a similar structure but $\Lambda^{\text{mode-0}%
}\approx3\Lambda^{\text{mode-1}}$. In SW reentries, the passage of each
activation front is associated with a profile of $A$ and $D$ holding one
(mode-0, panel A) or multiple (mode-1, panel C) maxima and minima over two
turns. For DW\ QP reentries, $L$ is twice the value\ for SW\ QP reentries,
\ such that mode-0 has one maximum and one minimum over one turn. (fig.
\ref{quasi_sust}A). Successive activations at each site alternate between long
and short $A$ and $D$ values, except at a number of nodes corresponding to the
boundaries from which the phase of the alternation is inverted. As illustrated
by the time-course of $T_{act}(x)$ and $T_{rep}(x)$(fig. \ref{quasi_sust},
right panels), the quasi-periodic nature of the propagation\ makes the
position of the extrema and of the nodes to drift slowly in the direction
inverse to the propagation of the activation fronts. QP reentry is thus
constituted by discordant alternans
\cite{Watanab2001_Me,Fox2002_Io,Echebar2002_In} with boundaries between short
and long APD moving around the loop.%

Period-1 SW reentry can be annihilated by an isolated stimulus applied in the
narrow vulnerable window in which the stimulation produces only a retrograde
front, corresponding to the well-known mechanism of unidirectional block
\cite{Starobi1994_Vu,Shaw1995_Th,Fei1996_As}. In a previous paper, we have
also described other modes of annihilation as well as different transient
dynamics that were induced by two successive stimuli \cite{Comtois2002_Re}.
These new modes\ of annihilation were compatible with experimental
observations, and relevant to antiarrhythmic pacing therapy
\cite{Mensour2000_In}. However, that study was restricted to a specific length
of the loop ($L_{crit}<L=25$ cm $<2L_{\min}$). In the following, we present a
systematic investigation of the outcomes of double stimuli applied on period-1
SW reentry as a function of the timing of the pulses and the length of the loop.

\subsection{Functional heterogeneity of refractoriness}

\begin{figure}
[ptb]
\begin{center}
\includegraphics[
height=5.8973in,
width=3.5456in
]%
{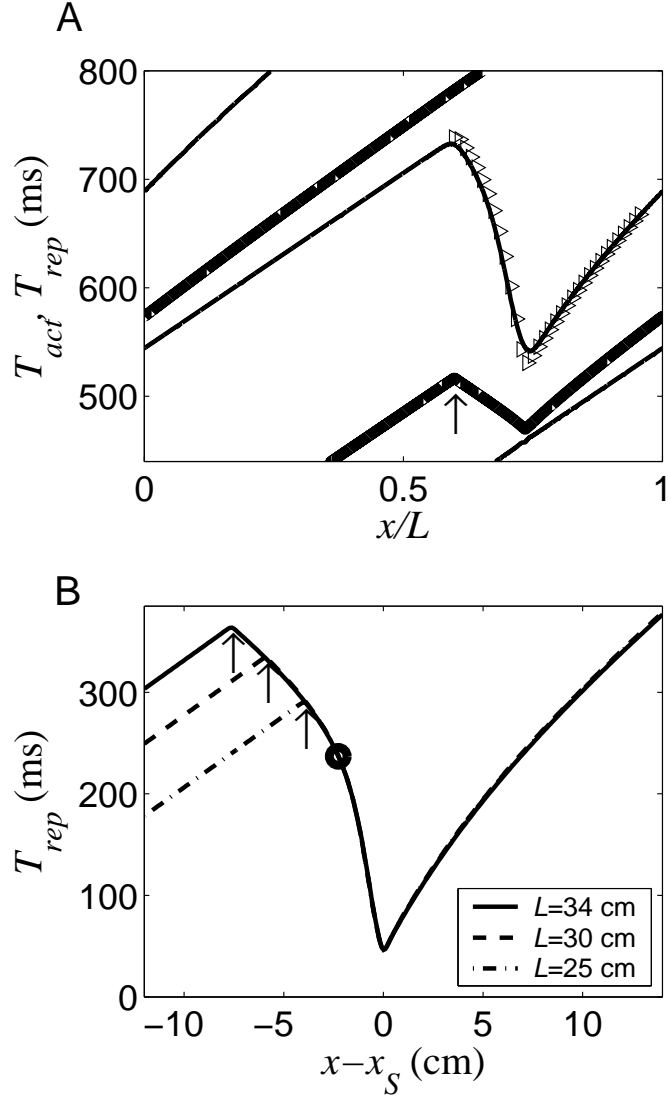}%
\caption{A) $T_{act}(x)$(thick line) and $T_{rep}(x)$ (thin line: IL model,
${\large \vartriangleright}$: ID model) following the application of a
stimulus $S_{1}$ at a site $x_{s}$ on a loop of 23 cm holding a period-1 SW
reentry. $S_{1}$ produces an antegrade front $A_{1}$ propagating to the right,
and a retrograde front $R_{1}$ that collides with the front at the reentry at
the site $x_{c}$ indicated by the arrow. B) $T_{rep}(x)$ around the
stimulation site $x_{S}$ for loops of \ 25, 30 and 34 cm, using the ID model.
$\Delta_{F-S1}$ was chosen such that $D(x_{S})=\Delta_{F-S1}-A_{F}(L)$ was the
same on all loops, $A_{F}(L)$ being the duration of the action potential of
the reentry for each case. $x_{c}$ ($\uparrow$) is shifted to the left for
longer $L$. The circle indicated the site where $\left\vert dT_{rep}%
/dx\right\vert =$ $c_{t,\max}$.}%
\label{fig_gradtrep}%
\end{center}
\end{figure}

Complex dynamics can be induced by a second stimulus $S_{2}$ thanks to the
asymmetric profile in $T_{rep}(x)$ left by the interaction of first stimulus
$S_{1}$ with the reentry activation front $F$. When $\Delta_{F-S1}$, the time
between the last passage of $F$ and the onset of $S_{1}$, is beyond the
vulnerable window, $S_{1}$ produces both a retrograde ($R_{1}$) and an
antegrade ($A_{1}$) activation front.\ As illustrated in fig.
\ref{fig_gradtrep}A, the key factor determining the dynamics that can be
induced by $S_{2}$ is the region located between the site of stimulation
($x_{S}$) and the site of the collision between $F$ and $R_{1}$ ($x_{c}$,
identified by the arrow in \ref{fig_gradtrep}A). $T_{rep}$ (thin line in fig.
\ref{fig_gradtrep}A) is minimum near $x_{S}$ and reaches its maximum at
$x_{c}$. The IL and ID models produce the same profile of $T_{rep}(x)$ (thin
lines and triangles, respectively, in \ref{fig_gradtrep}A), showing that the
ID model that was initially developed to describe sustained reentry also
provides an appropriate low-dimensional representation of the dynamics when
stimulations are applied. The location of $x_{c}$ as well as the profile of
$T_{rep}(x)$ around $x_{S}$ depend on both $\Delta_{F-S1}$ and $L$. Figure
\ref{fig_gradtrep}B shows the profile of $T_{rep}(x)$ obtained from loops of
different lengths stimulated at the same diastolic interval $D_{S1}%
(x_{s})=\Delta_{F-S1}(L)-A_{F}(L)$, where $A_{F}(L)$ is the action potential
duration of the stable reentry for each $L$. The position of $x_{c}$ is
shifted to the left (arrows in \ref{fig_gradtrep}B) because the collision is
delayed on longer loops. However all the loops have the same invariant profile
of $T_{rep}$ in the time and space interval that they share before the
collision. If $S_{1}$ is applied at larger $D_{S1}$ value, the distance between
$x_{S}$ and $x_{c}$ is shortened, $A(x_{S})$ and $T_{rep}(x_{S})$ are
increased, such that the extent and depth of the cusp in $T_{rep}$ around
$x_{S}$ are diminished.

\

\subsection{Initiating a second antegrade propagation\label{sec_creation_2fr}}

\begin{figure}
[ptb]
\begin{center}
\includegraphics[
height=3.9504in,
width=4.2969in
]%
{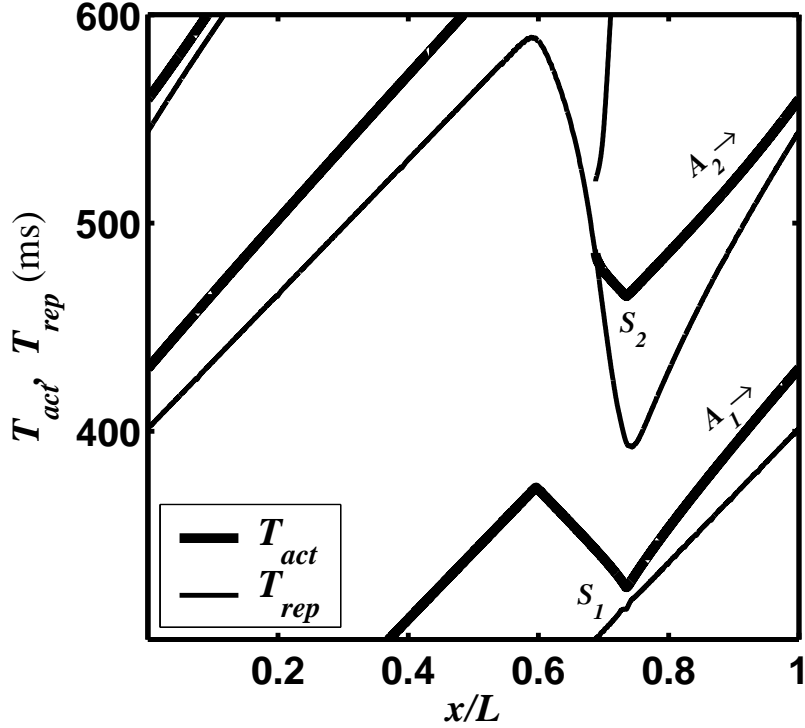}%
\caption{IL model, on a loop of 23 cm: $S_{2}$ creates both an antegrade
($A_{2}$) and a retrograde front $R_{2}.R_{2}$ is blocked between $x_{S}$ and
$x_{c}$ shortly after its creation in the refractory tail left by the
retrograde front $R_{1}$produces by $S_{1}$.}%
\label{blockR2}%
\end{center}
\end{figure}

The spatial profile of $T_{rep}$ for short $\Delta_{F-S1}$ is asymmetrical,
with a sharp gradient between $x_{S}$ and $x_{C}$, and a more gradual increase
at the right of $x_{S}$. Owing to this asymmetry, the outcome of $S_{2}$
depends on $\Delta_{S1-S2}$, the time interval between the onset of the two
stimuli. Figure \ref{blockR2} illustrates a case in which $S_{2}$ is applied
after the collision of $F$ and $R_{1}$, at an instant where $A_{1}$, the
antegrade front created by $S_{1}$, still has not reached $x_{c}$. $S_{2}$
creates both an antegrade ($A_{2}$) and a retrograde ($R_{2}$) front, but
$R_{2}$ is blocked between $x_{S}$ and $x_{c}$. Thereupon, the system is left
with two antegrade fronts ($A_{1}$ and $A_{2}$). This occurs as long as
$R_{2}$ does not propagate beyond $x_{c}$ to collide with $A_{1}$, in which
case $A_{2}$ is left alone to perpetuate the reentry. This is an alternate
scenario of unidirectional block that creates a propagating wave in the same
direction as $F$ and $A_{1}$.%

\begin{figure}
[ptb]
\begin{center}
\includegraphics[
height=3.9779in,
width=4.1803in
]%
{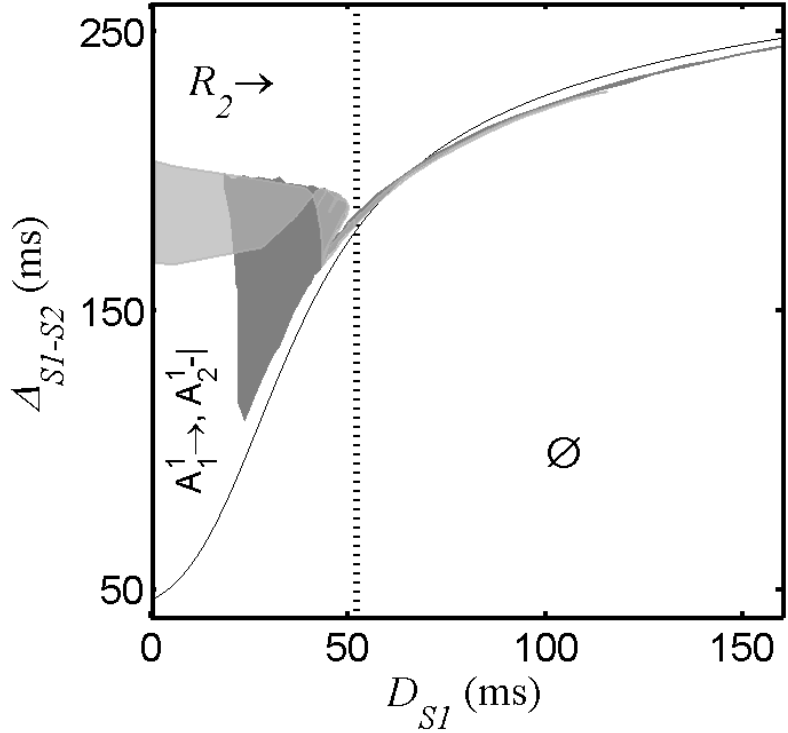}%
\caption{IL model: Outcome of the $S_{1}-S_{2}$ protocol applied on a SW
period-1 reentry in the [$D_{S1},\Delta_{S1-S2}$] plane for two loops with
$L=25$ \ and $34$ cm. The two loops have a common lower Region ($\varnothing$)
where \ $S_{2}$ is applied in the refractory period, and upper region
(\ $R_{2}\rightarrow$) where $R_{2}$ propagates beyond $x_{c}$. In the area in
between, $R_{2}$ is blocked and complex dynamics may occur. They take place in
a specific subset for each $L$: $L=25$ cm, ligth gray, $L=34$ cm, dark gray.}%
\label{fig_map_super}%
\end{center}
\end{figure}

All the complex dynamics occur in the range of $\Delta_{S1-S2}$ for which
$R_{2}$ is blocked between $x_{S}$ and $x_{c}$ whereas $A_{2}$ propagates.
This depends on the profile of $T_{rep}\left(  x\right)  $ left by $S_{1}$,
which was shown to be an invariant function of $D_{S1}$ in fig.
\ref{fig_gradtrep}B. Figure \ref{fig_map_super} shows the global
characteristics of the dynamics in the [$D_{S1,}\Delta_{S1-S2}$ ] parameters
plane for two values of $L$ ($L=25cm<2L_{\min}$ and $2L_{\min}^{\text{mode-1}%
}<L=34$ cm $<2L_{crit}$). The parameter plane can be divided in three areas.
In the region labeled $\varnothing$ at low $\Delta_{S1-S2}$ values,
$S_{2\text{ }}$is applied during the refractory period and does not produce a
response. In the upper region, labeled "$R_{2}\rightarrow$", $R_{2}$
propagates beyond $x_{c}$, collides with $A_{1}$, and $A_{2}$ is left alone to
maintain the reentry. In the middle area, $R_{2}$ is blocked between $x_{c}$
and $x_{s}$and complex dynamics may occur. The phase plane area in which
$R_{2}$ is blocked, for $D_{S1}$, between $0$ and $\sim50$ ms (dotted vertical
line), is the same for the two $L$. The specific subsets in which complex
dynamics occurs (represented by different shaded areas in fig.
\ref{fig_map_super}) change with $L$, and are discussed later.

The article is focused on the area where $R_{2}$ is blocked, between
$D_{S1}=0$ and $\approx50$ ms. The lower limit of the area\ coincides with
$A\left(  D_{S1}\right)  $ and is close to the action potential duration
restitution curve $A^{SP}\left(  D\right)  $. The upper limit is nearly a
constant, around $\Delta_{S1-S2,\max}\approx204$ ms. Appendix \ref{appendix_A}
shows that this $\Delta_{S1-S2}$ upper limit\ is set by the locus between
$x_{s}$ and $x_{c}$\ where $|dT_{rep}\left(  x\right)  /dx|$ is equal to the
maximum conduction time. This locus, indicated by a circle for the specific
$D_{S1}$ illustrated in \ref{fig_gradtrep}B, does not depend on $L$, which
explains why the upper limit for the block of $R_{2}$ is identical in the two
loops. $|dT_{rep}\left(  x\right)  /dx|$ remains everywhere below the maximum
conduction time if $D_{S1}$ is too long, explaining why the area of complex
dynamics disappears beyond $D_{S1}$ $\approx50$ ms. It is also demonstrated in
Appendix \ref{appendix_A} that the maximum slope of the $A^{SP}\left(
D\right)  $ function must be greater than $\sim1$ to allow the block of
$R_{2}$. The same condition that controls the stability of the period-1
reentry \cite{Courtem1996_A,Cytrynb2002_St,comtoispre2003} thus determines if
complex dynamics can be induced by $S_{2}$.%

\subsection{Interactions between the two antegrade propagating fronts}

Once $A_{2}$ has started to propagate and $R_{2}$ has been blocked, there are
4 possible outcomes:

1) $A_{1}$ is blocked and $A_{2}$ perpetuates the SW reentry,

2) $A_{2}$ is blocked, and $A_{1}$ maintains the SW reentry,

3) $A_{1}$ and $A_{2}$ are blocked, and reentry is annihilated

4) neither $A_{1}$ nor $A_{2}$ are blocked, and there is a transition to DW
reentry. As seen in table \ref{table_attract}, this last option can only occur
if $L\geq2L_{\min}$.

The next section discusses the cases 3) and 4) in which the system does not
return to the original SW period-1 reentry.

\subsubsection{Reentry annihilation by alternans amplification ($A_{1}$ and
$A_{2}$ are blocked)}

\begin{figure}
[ptb]
\begin{center}
\includegraphics[
height=5.9059in,
width=3.5653in
]%
{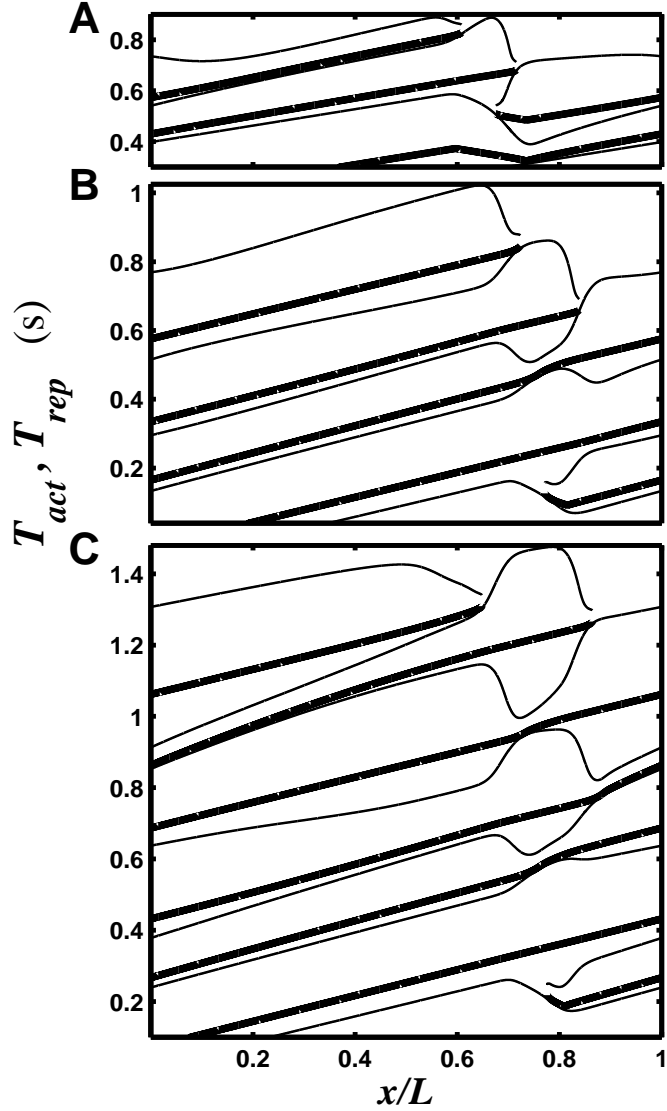}%
\caption{IL model: profile of $T_{act\text{ }}$(thick line) and $T_{rep}$
(thin line) for the three different scenarios of annihilation caused alternans
amplification model in the ring. A) $L=23$ cm, $D_{S1}=10$ ms, $\Delta
_{S1-S2}=160$ ms, [$A_{1}^{1}\dashv,A_{2}^{1}\dashv$], B) $L=25$ cm,
$D_{S1}=47$ ms, $\Delta_{S1-S2}=190.4$ ms, [$A_{1}^{2}\dashv,A_{2}^{2}\dashv
$], C) $L=25$ cm, $D_{S1}=47$ ms, $\Delta_{S1-S2}=185.1$ ms [$A_{1}^{3}%
\dashv,A_{2}^{3}\dashv$]. $S_{1}$ and $S_{2}$ are pictured in panel A, but
only $S_{2}$ is shown in panel B and C.}%
\label{fig_bloc_aa}%
\end{center}
\end{figure}

The three new scenarios of reentry annihilation reported in
\cite{Comtois2002_Re} are shown in fig. \ref{fig_bloc_aa}. These scenarios of
termination differ in the number of revolution made by the $A_{1}$ and $A_{2}%
$activation fronts before they are blocked. Accordingly, we introduce the
notation $A_{i}^{m}\dashv$ , meaning that the front $A_{i}=\{1,2\}$\ is
blocked ($\dashv$) after $m$ turns around the loop. Let consider the simplest
case ([$A_{1}^{1}\dashv,A_{2}^{1}\dashv$], fig. \ref{fig_bloc_aa}A) in which
both $A_{1}$ and $A_{2}$ are blocked after one rotation. $A_{1}$ is blocked
first near $x_{S}$, when it reaches the refractory tail left by $A_{2}$ and
$R_{2}$. This occurs because $A_{1}$ has already completed a fraction of its
rotation when $S_{2}$ is applied. As a consequence, $A_{1}$ comes back to
reactivate the sites near $x_{S}$ after a time interval much shorter than the
period of rotation $T$ of the stable reentry. The block of $A_{2}$ takes place
between $x_{c}$ and the site where $R_{2}$ was blocked. When $A_{1}$ travels
in this zone, the region has last been excited by $R_{1}$, and a time interval
longer than $T$ has elapsed since this last excitation. Besides, both the
action potential and $T_{rep}$ associated to $R_{1}$ were short since $S_{1}$ 
was premature. As a consequence, $A_{1}$ produces action potentials that are
longer than those of the stable reentry. Since the time between the passage of
$A_{1}$ and the return of $A_{2}$ is also shorter than $T$, $A_{2}$ is
blocked. The block of $A_{2}$ thus results from a process of amplified
alternation in a region between $x_{S}$ and $x_{c}$. The premature $R_{1}$
that creates short action potential is followed by the late $A_{1}$ generating
long action potential.

In the two other scenarios, $A_{1}$ nor $A_{2}$ are blocked after either 2 or
3 turns ([$A_{1}^{2}\dashv,A_{2}^{2}\dashv$], [$A_{1}^{3}\dashv,A_{2}%
^{3}\dashv$] in fig. \ref{fig_bloc_aa}B-C). Each passage of $A_{1}$ leaves
around $x_{S}$ a convex profile of $T_{rep}$, which is turned to a concave
profile by the subsequent passage of $A_{2}$. The panel B and C illustrates
the process of alternans amplification. Alternans amplification may end up by
the blockade of one of the front, in which case propagation reverts to SW
reentry. It may saturated, leading to DW reentry, or may lead to annihilation.
The next section explores the conditions for annihilation.%

\paragraph{The scenarios of annihilation by alternans amplification occurs
over limited range of $L$}

\begin{figure}
[h]
\begin{center}
\includegraphics[
height=6.2361in,
width=3.8972in
]%
{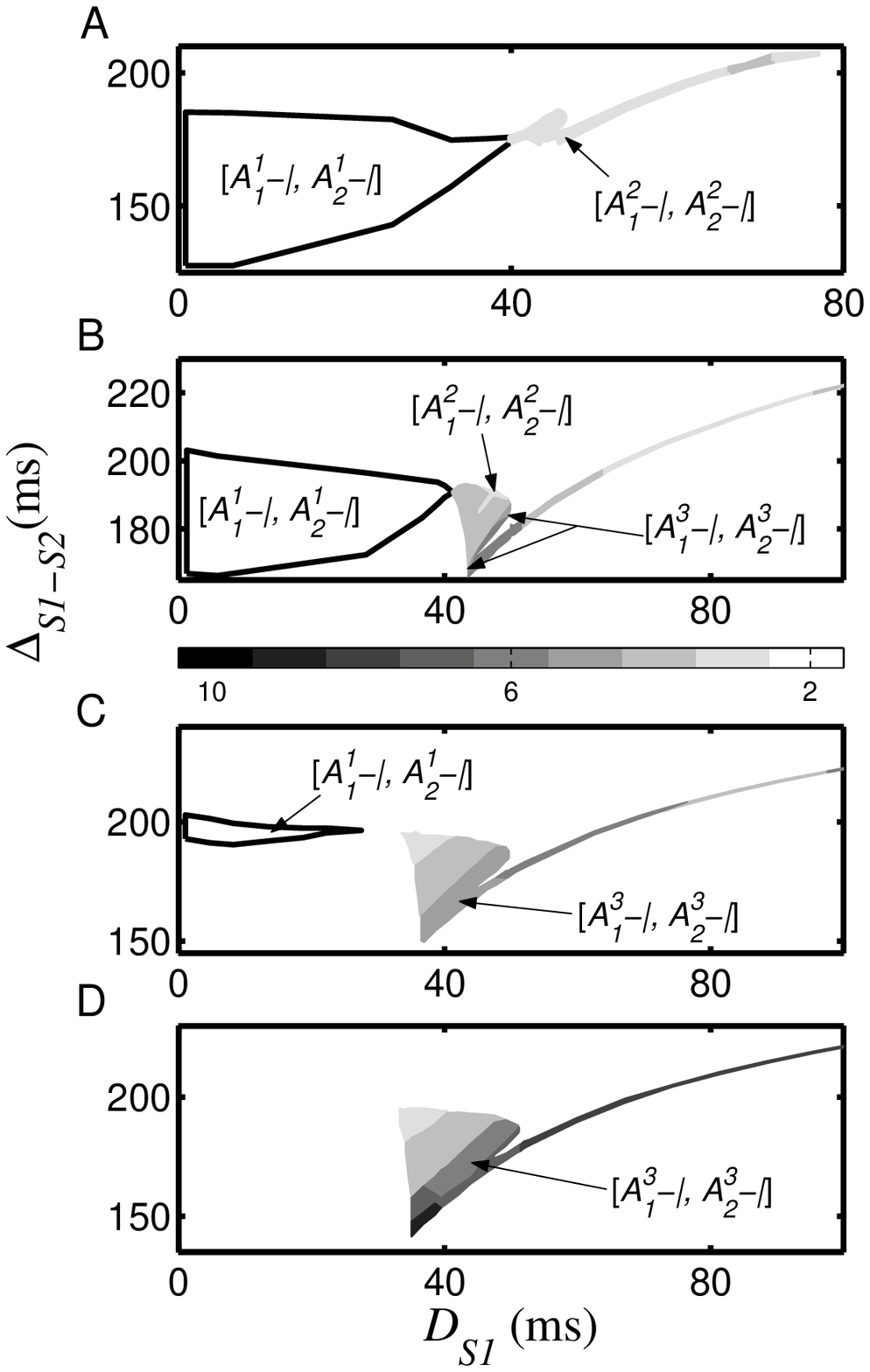}%
\caption{IL\ model: Area with complex dynamics and alternans amplification in
the ($D_{S1},$ $\Delta_{S1-S2})$ plane with A to D) $L=20.5$, $25.0$, $27.5$,
$28.5$ cm. \ The gray scale indicates the number of turns during which $A_{1}$
and $A_{2}$ persist together. The regions with different annihilation are
specified accordingly. Otherwise, one front ends up being blocked and
propagation reverts to period-1 SW reentry.}%
\label{fig_map_single}%
\end{center}
\end{figure}

The conditions leading to each type of alternans amplification annihilation
vary with $L$, and none exists for $L$
$>$%
$\sim30.5$ cm. The four panels of fig. \ref{fig_map_single} picture the extent
of the different zones of annihilation for $L$ ranging from $L_{crit}=19.5%
\operatorname{cm}%
$, the minimum $L$ with stable period-1 SW reentry, to the limiting
$L\approx30.5$ cm value. At short $L$ ($L=20.5$ $%
\operatorname{cm}%
$, fig. \ref{fig_map_single}A), there is a large [$D_{S1}$, $\Delta_{S1-S2}$]
area with [$A_{1}^{1}\dashv,A_{2}^{1}\dashv$] block located at low $D_{S1}$
values, and a small adjacent area of [$A_{1}^{2}\dashv,A_{2}^{2}\dashv$]
\ block. The zone of [$A_{1}^{2}\dashv,A_{2}^{2}\dashv$] \ always remains
minimal; and it is the first to disappear at\ $L\approx28.5$ cm. The zone of
[$A_{1}^{3}\dashv,A_{2}^{3}\dashv$] appears at intermediate $L$, expands, and
is the last to disappear.%

\paragraph{Disappearance of the\ [$A_{1}^{1}\dashv,A_{2}^{1}\dashv$]
annihilation}

Taken together, the maps of fig. \ref{fig_map_super} and fig.
\ref{fig_map_single} A-C shows that the area of [$A_{1}^{1}\dashv,A_{2}%
^{1}\dashv$] is embedded in the larger region, invariant with respect to $L$,
in which $R_{2}$ is blocked between $x_{c}$ and $x_{S}$. The zone of
[$A_{1}^{1}\dashv,A_{2}^{1}\dashv$] is located at intermediate $\Delta
_{S1-S2}$, just over the area, labeled [$A_{1}^{1}\rightarrow,A_{2}^{1}\dashv
$]\ (fig. \ref{fig_map_super}), in which $A_{1}$can propagated beyond $x_{S}$,
but $A_{2}$ is stopped by the refractory tail left by $A_{1}$.\ As $L$ is
increased, the lower boundary of [$A_{1}^{1}\dashv,A_{2}^{1}\dashv$] is
lifted,\ thus diminishing the [$A_{1}^{1}\dashv,A_{2}^{1}\dashv$] area until
it disappears completely. The lost of [$A_{1}^{1}\dashv,A_{2}^{1}\dashv$]
annihilation is thus caused by the inability of the system to block $A_{1}$ at
its first return.

Appendix \ref{appendix_B} provides the conditions needed for $A_{1}$ to block
in the tail of $A_{2}$ and proves that there is a limiting $L$ beyond which
this cannot happen. To summarize: 1) Increasing $\Delta_{S1-S2}$ produces
longer action potential for $S_{2}$, delays $T_{rep}(x_{S})$, and thus
augments the likelihood of $A_{1}$ to be blocked. 2) However, the increase of
$\Delta_{S1-S2}$ and $T_{rep}(x_{S})$ are bounded by the condition of $R_{2}$
being blocked between $x_{S}$ and $x_{c}$, and these limiting values are
independent of $L$. 3) Since increasing $L$ delays the return of $A_{1}$,
there is a length $L$ from which the return of $A_{1}$ always occurs after the
limiting $T_{rep}(x_{S})$ value.

\paragraph{Opening the way to more than one rotation for both fronts}

Figure \ref{fig_map_single} shows that, for each value of $L$, the zone of
[$A_{1}^{1}\dashv,A_{2}^{1}\dashv$] \ and [$A_{1}^{2,3}\dashv,A_{2}%
^{2,3}\dashv$] blocks are disjoint, being respectively located at low and high
$D_{S1}$. As seen in fig \ \ref{fig_map_single}D,\ the zones of [$A_{1}%
^{2,3}\dashv,A_{2}^{2,3}\dashv$] remain located at large $D_{S1}$ even at
values of\ $L$ for which [$A_{1}^{1}\dashv,A_{2}^{1}\dashv$] \ does not exists
anymore. However, enlarging $L$ also leads to the appearance and extension of
areas in which $A_{1}$ and $A_{2}$ persist together for an increasing number
of turns before one of the front is blocked (up to 7 activations in
\ref{fig_map_single}D). As $L$ is increased, the areas of the parameters plane
associated to these other forms of transient complex dynamics extend toward
low\ $D_{S1}$, to finally cover all the range from $D_{S1}=0$ to $50$ ms when
$L=37.5$ cm, a value close to $2L_{crit}$ from which DW reentry becomes
possible. The areas of the parameters plane associated to these behaviors with
coexistence of \ $A_{1}$ and $A_{2}$ for multiple turns form a sequence of
contiguous parallel tongues.

All the higher modes of block and complex dynamics require that both $A_{1}$
and $A_{2}$ travel beyond $x_{s}$, at least at their first return. Conditions
for the block of $A_{2}$ are derived in Appendix \ref{appendix_C} based on an
approximation using the $A^{SP}\left(  D\right)  $ equation and rules for
$A_{1}$ and $A_{2}$ return cycles. It shows that the block of $A_{2}$ depends
on the balance between the return cycle of $A_{1}$ and $A_{2}$.

\subsubsection{Transition to double-wave reentry (neither $A_{1}$ nor $A_{2}$
is blocked)\label{res_trans_qp}}

\begin{figure}
[tbh]
\begin{center}
\includegraphics[
height=4.7275in,
width=3.2583in
]%
{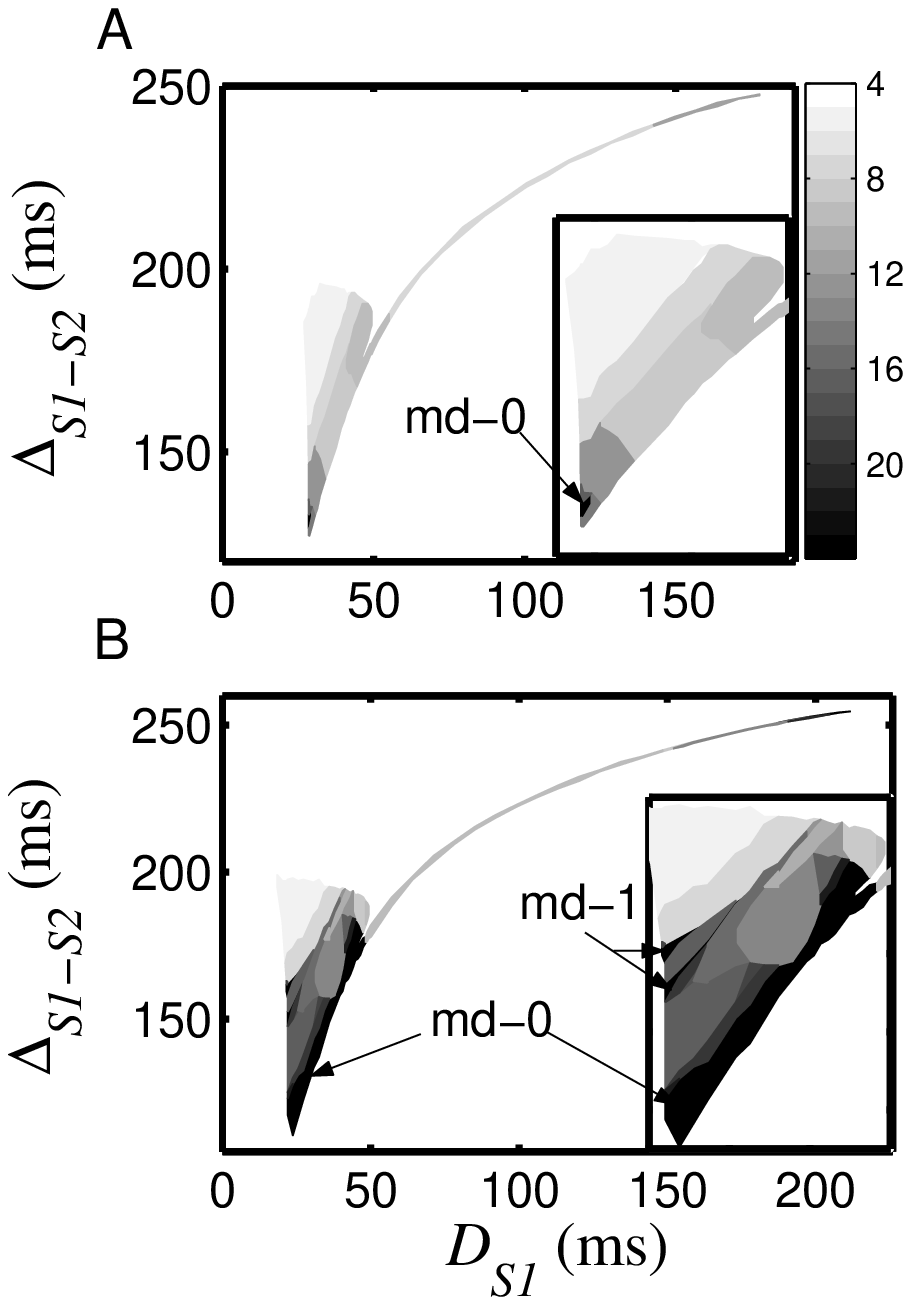}%
\caption{IL model: area of complex dynamics in the $D_{S1}$ vs. $\Delta
_{S1-S2}$ plane for $L=31$ cm (panel A) and $L=34$ cm (panel B). The gray code
indicates the number of turns during which $A_{1}$ and $A_{2}$ persist
together. Arrows pinpoint the zones with transition to mode-0 and mode-1 DW QP
dynamics.}%
\label{map_6200}%
\end{center}
\end{figure}

In both the ID and IL model, annihilation by alternans amplification is
impossible from $L$ $\geq$ 30.5 cm. In fact, for 31 $>L\ \geq$ 30.5 cm, only
transient complex dynamics with a final return to period-1 SW reentry are
observed. However, the maximum number of turns with $A_{1}$ and $A_{2}$
co-traveling grows, just as the number of\ tongues in the [$D_{S1}%
,\Delta_{S1-S2}$ ] plane associated to different numbers of turns during which
the two fronts coexist. Each new tongue appears at low $D_{S1}$ and
$\Delta_{S1-S2}$ values, and expands as $L$ is further increased.\ Finally, a
first transition to QP mode-0 DW reentry is detected at $L=31$ cm (fig.
\ref{map_6200}A). Transition to DW reentry thus appears as the asymptotic
limit of the prolongation of the transient propagation with two fronts.
However, transition to DW reentry begins much beyond $L=2L_{\min}$ $=25.6$ cm,
the value at which sustained mode-0 QP DW reentry starts to exist. In fact, at
$L=31$ cm, the system is rather in the range of $L$ for which both DW mode-0
and mode-1 solutions coexists.%

For longer $L$, (fig. \ref{map_6200}B) the region with transition to mode-0
DW reentry expands, as it was the case for all the zones associated to two
fronts transient propagation created at shorter $L$. Transition to mode-1 DW
reentry also appears. However, the transition to mode-1 appears at low
$D_{S1}$, but at two disjoint intermediate $\Delta_{S1-S2}$ values, giving rise
to two separated areas.

For still longer $L$, the areas of mode-0 and mode-1 transition enlarge until
covering, from $L=37.5$ cm, all the area where $R_{2}$ is blocked. Hence, from
this value of $L,$transition to DW reentry occurs in all instances where
$R_{2}$ is blocked. Finally, beyond $2L_{crit}=39$ cm, all transitions go to
period-1 DW reentry, which is the only DW solution that remains.

\subsubsection{Selecting the mode of QP reentry}

Increasing the length of the pathway correlates with an increase in the
complexity of the dynamics and to the transition to sustained DW reentry. The
results presented in fig. \ref{map_6200} show that changing the timing of the
stimuli when $2L_{\min}<L<2L_{crit}$ can induce the transition either to the
mode-0 or mode-1 DW QP reentry. The IL model was simulated with $L=37.5$ cm
($2L_{\min}\leq L<2L_{crit}$ but near $2L_{crit}$) to circumscribe the basin
of attraction in the parameter space associated to each DW QP reentry. The
large area in which two antegrade fronts are created ($D_{S1}<50$ ms) is
separated between two regions, with lower $\Delta_{S1-S2}$ converging to
mode-0, and higher $\Delta_{S1-S2}$ to mode-1. In this last section, we
compare the\ transient dynamics leading to each of two modes of DW QP reentry.

\begin{figure}
[tbh]
\begin{center}
\includegraphics[
height=4.7026in,
width=3.4195in
]%
{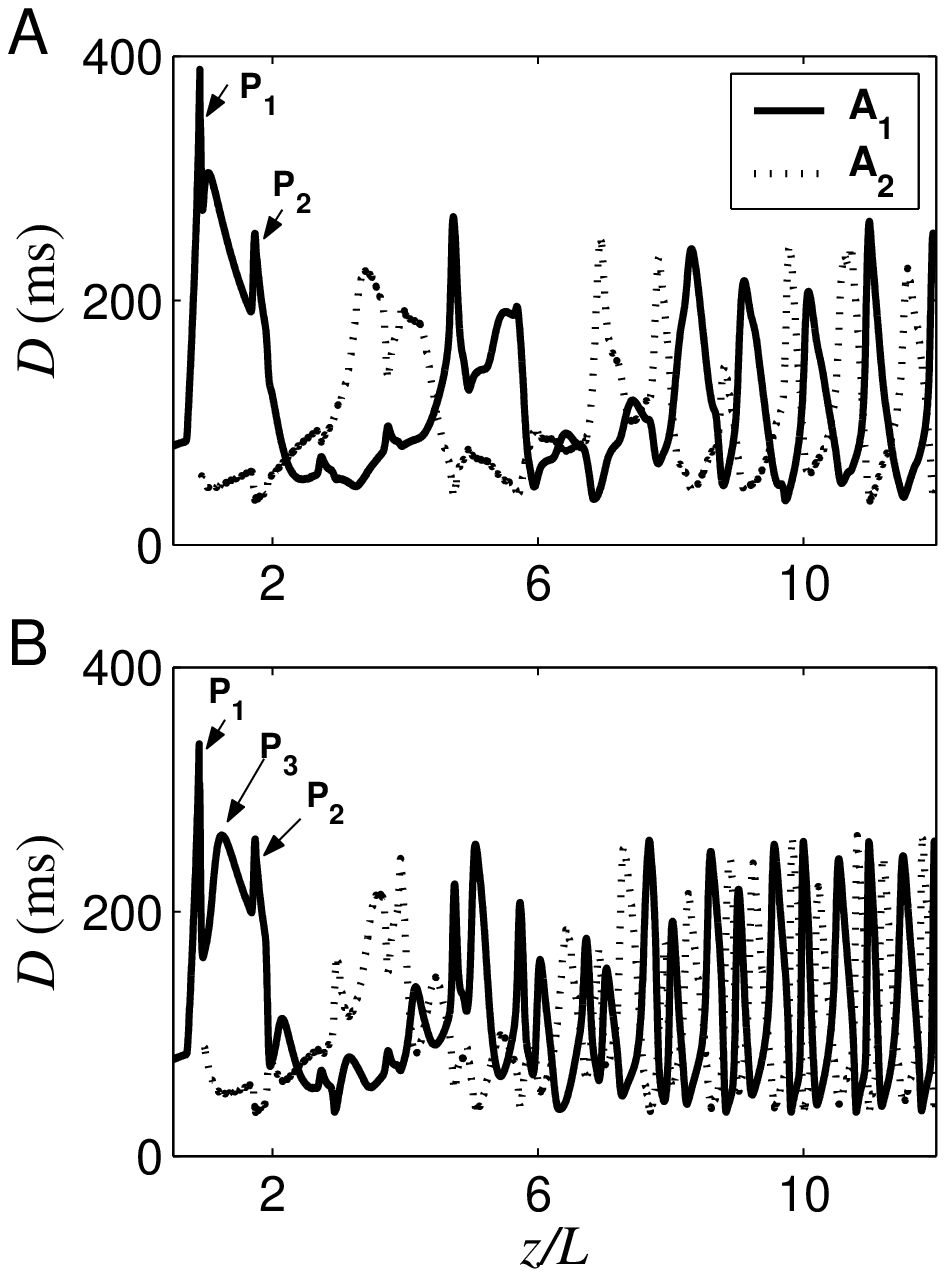}%
\caption{The spatial variation of $D$ as a function of the travelled distance
for the transition to A) mode-0 and B) mode-1 double-wave QP reentry.}%
\label{fig_di_quasi}%
\end{center}
\end{figure}

Figure \ref{fig_di_quasi} \ shows the spatial profile of $D$ associated to
$A_{1}$ and $A_{2}$ for two cases converging respectively to mode-0 (panel A)
and mode-1(panel B). In this representation,$\ $the passage of $A_{2}$
($D_{A2}\left(  z\right)  $, dotted line) at each location is followed by that
of $A_{1}$ ($D_{A1}\left(  z\right)  ,$full line). At first, $A_{2}$
propagates with a short $D$, producing brief action potentials everywhere
along the loop. As a consequence, $A_{1}$ travelling afterwards meets long
$D\left(  z\right)  $. $D_{A1}\left(  z\right)  $ has its first maximum
(P$_{1}$) when it travels between $x_{c}$ and the point where $R_{2}$was
blocked, such that $D_{A2}$ has a minimum at the same location from which it
increases slowly until its next return in the same region. From there,
$D_{A2}\left(  z\right)  $ starts to alternate between long and short values
with a spatial period close to 2 turns, and $D_{A1}\left(  z\right)  $ follows
a complementary profile. This first phase of the propagation, lasting for
approximately 8 turns, can be labeled as concordant alternans since $D_{A2}$
and $D_{A1}$ have alternating values and that each remains short or long for
at least one complete rotation. This pattern does not correspond neither to
mode-0 nor mode-1, since both stabilized DW QP solutions have a wavelength
less than $L$.

However, from the beginning ($k=1,2$), $D_{A1}$ already shows a second spatial
oscillation in $D$ that is superimposed to the concordant alternans. The
structure of this oscillation, which has a wavelength close to one turn, makes
the difference between the cases converging to mode-0 and mode-1. In the
former case, the oscillation embeds\ two peaks \{P$_{1}$, P$_{2}$\} ($D_{A1}$
for mode-0, continuous line in fig. \ref{fig_di_quasi}A), while in the later
case, it has three peaks \{P$_{1}$, P$_{2}$, P$_{3}$\} ($D_{A1}$ for mode-1,
continuous line in fig.\ \ref{fig_di_quasi}B). These superimposed spatial
variations persist while the concordant alternans dissipate. During this
process, the position of the peaks does not change much. As the amplitude of
the concordant alternans decreases, the respective height of the discordant
alternans increases up to a point where the boundary with large gradient in
$D$ begins to move around the loop due to the quasiperiodic nature of the
propagation. This travelling mechanism is akin to the propagation of paced
discordant alternans on a cable of cardiac
tissue\cite{Watanab2001_Me,Fox2002_Io,Echebar2002_In}.%

The main difference between the transition to mode-0 and mode-1 is the
presence of the third peak P$_{3}$ in $D_{A1}$. The P$_{3}$ peak of $D_{A1}$
is induced by the increase of $\Delta_{S1-S2}$. On one hand, a larger
$\Delta_{S1-S2}$ produces a longer $T_{rep}$ such that $A_{1}$ meets a lower
$D$ and the amplitude of P$_{1}$ is reduced. But a larger $\Delta_{S1-S2}$
also implies that $A_{2}$ travels faster, comes back sooner to the stimulation
sites, and set the stage for a new maximum in the $D_{A1}$ profile.

Of course the transition from two to three peaks is continuous process since a
similar variation of $D$ with less amplitude already exists with the
transition to mode-0 in fig. \ref{fig_di_quasi}A. It means that there must be
a minimal spatial profile that corresponds to the boundary between the two
basins of attraction (the transition to either mode-0 or mode-1).

\section{Discussion}

Alternans amplification, leading either to reentry annihilation or transition
to DW reentry, can exist if the slope of the APD restitution curve is larger
than one. Hence, the condition on the slope of the APD restitution curve that is
mandatory for the existence of\ sustained QP reentry\cite{comtoispre2003} also
enables double-pulse stimulation to produce a new mode of unidirectional
block, in which only an antegrade front propagates away from the stimulation
site. On any closed circuit with two activation fronts travelling in opposite
direction, this opens a large spatio-temporal window in\ which an ectopic
focus or an external source firing twice can start a reentry.\ This is
consistent with the use of burst pacing as a standard experimental and
clinical procedure to start tachycardia
\cite{Vinet1996_Cy,Cao1999_Sp,Helie2000_Cy,Nattel2000_Ba}. In a loop already
holding a SW reentry,\ the $S_{1}-S_{2}$ protocol can induced the simultaneous
propagation of\ two antegrade fronts whose final outcome depends on the timing
of the stimuli and the length of the loop. The transient or persistent
coexistence of two antegrade fronts is a new type of dynamics in which the
effects of the stimuli cannot be represent as perturbations of a limit-cycle,
as it has been done for models in which the steep slope criteria was not
fulfill \cite{Nomura1996_En,Glass2002_Pr}.

In clinical and experimental investigations, stimulations are currently used
to either study the characteristics of reentry circuits through resetting or
to stop the tachycardias \cite{Fei1996_As,Jalil2003_Ex}. Most often,
stimulations are applied at one site and propagation is assessed through one
or a few recording electrodes. We may consider the dynamics that will be
observed \ with this setting for each of the three scenarios of annihilation
by alternans amplification. Since the distance traveled by $R_{1}$ from
$x_{s}$ to $x_{c}$\ usually covers a small portion of the reentry pathway,
electrodes are much more likely to be positioned in the complementary portion
of the circuit. For [$A_{1}^{1}\dashv,A_{2}^{1}\dashv$], \ such electrodes
will detect the last passage of $F$ before the collision with $R_{1}$, then
$A_{1\text{ }}$and $A_{2}$. The detection of $A_{1\text{ }}$and $A_{2}$ would
clearly exclude classical unidirectional block being responsible for the
annihilation. In fact, this modeling study was initiated after a set of
experimental and clinical studies on flutter using multichannel (4 to 8
channels) recordings \cite{Mensour2000_In}. In these, cases of annihilation
were reported in which $F$, $A_{1\text{ }}$, $A_{2}$ were detected, and in
which the propagation of $A_{1\text{ }}$and $A_{2}$ was blocked in the segment
of the circuit were the collision of $F$ and $A_{1\text{ }}$was presumed to
have occurred. This scenario, that was called collision block, is consistent
with the [$A_{1}^{1}\dashv,A_{2}^{1}\dashv$] block by alternans amplification
\cite{Comtois2002_Re}. In the case of [$A_{1}^{2}\dashv,A_{2}^{2}\dashv$]
block, $A_{1}$ and $A_{2}$ would be detected twice (see \ref{fig_bloc_aa}B).
Since the time intervals from each $A_{1}$ to the next $A_{2}$ are longer than
that between $A_{2}$ and the following $A_{1}$, the sequence of the
differences between activation times alternates around a value shorter than
the period of the original reentry, the block occurring after the longer
interval of the time series. For [$A_{1}^{3}\dashv,A_{2}^{3}\dashv$], each
front is seen thrice, with similar oscillation in the time series of \ the
difference. However, the structure of oscillation of the cycle as well as the
last value before annihilation depend on the position of the electrode in the
circuit. In a protocol for annihilation by unidirectional block, the detection
of $A_{1\text{ }}$and $A_{2}$ would indicate that the stimuli were beyond the
vulnerable window and would trigger the application of a new stimulus at
shorter coupling interval. However, this could rather reinitiate a reentry
bound to stop by alternans amplification.

To allow the unidirectional block of $R_{2}$, $S_{1}$ must be applied early
beyond the vulnerable window, in the portion of the excitable gap referred as
partially refractory by electrophysiologists
\cite{DellaBe1991_Us,Heisel1997_Fa}. The prematurity of\ $S_{1}$, in
conjunction with the high slope of the APD restitution curve and of the
$\theta$ dispersion curve, create a concave asymmetrical profile of
$T_{rep}(x)$ around $x_{s}$. For a given $S_{1}$, the $\Delta_{S1-S2}$
intervals for which $R_{2}$ is blocked depends on the local dynamics around
$x_{s}$ that is again determined by the APD restitution and the dispersion of
$\theta$. In fact, when the parameters $D_{S1}$ is used to describe the timing
of the first stimulus, the [$D_{S1},\Delta_{S1-S2}$] range to get a block of
$R_{2}$ becomes independent of $L$.

The pivotal role of the $A^{SP}(D)$ and $\theta(D)$functions is further
confirmed by the capacity of the ID model to reproduce the dynamics of the
ionic model. However, in order to avoid discontinuity in $T_{rep}(x)$ in the
region where $R_{2}$ is blocked and get a correct representation of the
dynamics, spatial averaging must be included in the computation of the spatial
profile of $T_{rep}(x)$. Originally, spatial averaging was added to the ID
model to reproduce the modulation of the repolarization by resistive coupling
in order to correct the shortcomings of the model regarding the details of the
bifurcation from periodic to QP propagation \cite{Vinet2000_Qu}.\ Including
spatial averaging becomes even more essential when stimulations and blocks
produce steep gradients in\ $T_{rep}(x)$ \cite{Comtois2002_Re}.

While the wide range of [$D_{S1},\Delta_{S1-S2}$]\ for which $R_{2}$ is
blocked does not depends on $L$, this is not the case for the subset of this
interval leading to reentry annihilation by alternans amplification. As shown
in fig. \ref{fig_map_single}, the interval of block is very large for $L\simeq
L_{crit}$ and decreases gradually until it disappears at $L\simeq2L_{\min
}^{\text{mode-1}}$. For the MBR model used herein, annihilation by alternans
amplification exists for $30.5>L>19.4$ cm, and it occurs on a range of
[$D_{S1},\Delta_{S1-S2}$] that remains much wider than the $\sim1$ ms standard
vulnerable window for most of this range of $L$. This is consistent with the
result of a clinical study in which dual pulses stimulation was found to be
four times more effective than single stimulus to stop monomorphic ventricular
tachycardia in man \cite{Almendr1986_An}. As $L$\ is increased, the
[$D_{S1},\Delta_{S1-S2}$] area of $R_{2}$ block also encloses a growing number
of tongues with increasingly prolong coexistence of $A_{1}$ and $A_{2}$, a
process that culminates in the appearance of zones of transition to DW
reentry.\ Then the zones with transition to DW reentry extend until they cover
completely the area in which $R_{2}$ is blocked. From $L\simeq2L_{crit},$
blocking $R_{2}$ yields automatically to sustained DW\ reentry. Hence,
transition to DW reentry does not necessarily imply a heterogeneous substrate
as it is been proposed elsewhere \cite{Brugada1991_On,Cheng1998_Ac}, but can
also be achieved in a homogeneous medium through the creation of a functional
heterogeneity by a limited number of electrical stimulations. Annihilation and
transition to DW occur on separated ranges of loop lengths. On shorter loop,
reentry annihilation is produced when alternans amplification reaches an
amplitude high enough to block $A_{1}$and $A_{2}$. On longer loop, the
distributed alternans saturate at an amplitude that still permits sustained
propagation. In our version of the BR model, the transition from SW QP to
period-1 reentry at $L_{crit}$ is supercritical. However, we have shown that,
for other sets of parameters, the bifurcation is subcritical, with bistability
between QP and period-1 reentry near $L_{crit}$ \cite{Vinet1999_Me}. In these
cases, it is possible that the two-stimulations protocol applied to the SW
period-1 reentry near $L_{crit}$ could induced a transition to SW QP
propagation, a phenomenon that was not possible with the instance of the MBR
model used in this paper.

Our results are consistent with different clinical and experimental
observations, and open the possibility to design more effective
anti-arrhythmic pacing strategies. However, the modeling studies must be
extended to more realistic representations to evaluate properly\ potential
applications. Preliminary results from an ongoing work on a two-dimensional
annulus show that block by alternans amplification can still be obtained on
this setting, but that additional scenarios are possible, including
termination through transient fibrillation that has also been observed in real
cardiac tissue. Tissue heterogeneity, either at the level of the ionic
properties or of the cells coupling, could also be important since termination
of reentry has been obtained in a one-dimensional loop model embedding a small
area of slower conduction but using an ionic model with minimal APD
restitution properties \cite{Sinha2002_Cr,Sinha2002_Te}. It remains to be seen
if annihilation based on alternans amplification would be amplified or reduced
by the inclusion of spatial inhomogeneity. Investigation will also have to be
extended to bidomain model in order to get a more proper representation of the
stimulus. The simplified representation used herein can be an acceptable
approximation for low amplitude stimuli or to mimic the effect spontaneous
firing of a group of cells. However it is known that current spread of the
stimulus depend on the properties of the external and internal medium
\cite{Lindblo2000_Ro,Keener2003_Th}. The MBR representation of the ionic
properties is also oversimplified. However, since most phenomena described in
this paper occurs in the few first beat after the stimulations and can be
explained from the APD restitution and speed dispersion, it is unlikely that
slow memory effects appearing in the dynamics of more complex model would
change the behaviour. Characterizing the APD restitution and speed dispersion
of the more complex model in the range of frequency of repetitive activity
associated to reentry should allow a prediction of the possible dynamics.

\section{Conclusion}

This work is a further illustration of the richness and diversity of the
dynamics that can results from the restitution of APD and dispersion of the
speed even in a simplified model of the tissue. It has revealed some
unexpected behaviours, like block by alternans amplification, which can be
much more prevalent than the mechanism of unidirectional block that are
usually assumed to be dominant. The low-dimensional model whose behaviour is
equivalent to the ionic model, provides a generic understanding of the
dynamics that can used a guideline to investigate the effects of future
complexification of the model.

\begin{acknowledgments}
This work was supported by grants from the Natural Sciences and Engineering
Research Council of Canada (AV), the \emph{Fonds Qu\'{e}b\'{e}cois de la
Recherche sur la Nature et les Technologies}(PC), as well as by the technical
and computer resources of the R\'{e}seau Qu\'{e}b\'{e}cois de Calcul de Haute Performance.
\end{acknowledgments}

\appendix

\section{The upper limit of the $\Delta_{S1-S2}$ interval for double-wave
creation\label{appendix_A}}

Three conditions must be fulfilled for the block to occur: 1) $R_{2}$, the
retrograde front created by $S_{2}$, must be blocked between the stimulation
site $x_{s}$and $x_{c},$ the locus of the collision between the reentry front
$F$ and retrograde front $R_{1}$ created by $S_{1}$; 2) afterward, $A_{1}$,
the antegrade front produced by $S_{1}$, must be blocked when it returns near
$x_{s}$; 3) finally, $A_{2}$, the antegrade front produced by $S_{2}$, must
also be blocked when it travels between $x_{s}$ and $x_{c}$ The following three
appendixes formulate the constraints associated with each of these conditions.
In all three appendixes, we consider that $F,A_{1}$and $A_{2}$ travels toward
increasing value of $x$, we use $T_{rep,F}(x_{s})$, the repolarization time
associated to the passage of the last\ activation front of the reentry $F$
at $x_{S}$, as the reference time $t=0$, and introduce a spatial coordinate
$y=\left(  x_{s}-x\right)  \geq0$ to follow the retrograde fronts $R_{1}$ and
$R_{2}$.

$R_{2}$ can propagate as long as its activation time $T_{act,R2}(x)$ is larger
than $T_{rep,R1}(x)$, the repolarization time associated to $R_{1}$. In the
limiting case $T_{act,R2}(x)=T_{rep,R1}(x)$, $D_{R2}(x)=T_{act,R2}%
(x)-T_{rep,R1}(x)=0$, and $R_{2}$ travels with the maximum conduction time
$-dT_{act,R2}/dx=c_{t}\left(  0\right)  $ =$c_{t,\max}$. When $R_{2}$ reaches a
point $x_{c}<x_{0}<x_{s}$ with $D_{R2}(x_{0})=0,$ it can continue to propagate
if $\left\vert dT_{rep,R1}(x_{0})/dx\right\vert $ $\leq c_{t,\max}$. The
limiting case for the propagation of $R_{2}$ is thus%

\begin{align}
T_{rep,R1}(x_{0})  &  =T_{act,R2}(x_{0})\label{cond_R2}\\
\left\vert \frac{dT_{rep,R1}(x_{0})}{dx}\right\vert  &  =c_{t,\max}\nonumber
\end{align}
\ We neglect the effect of coupling in the calculation of $T_{rep,R1}$ to obtain%

\begin{equation}
T_{rep,R1}\left(  y\right)  \simeq T_{act,R1}(y)+A^{SP}\left(  D_{R1}\left(
y\right)  \right)  \label{cond_R2s}%
\end{equation}
where $D_{R1}\left(  y\right)  =$ $T_{act,R1}(y)-T_{rep,F}(y)$ is the
diastolic interval associated to the propagation of $R_{1}$. From our choice
of reference time and the definition of $y$, $T_{act,R1}(0)=D_{S1}$. In the
MBR model, the $c_{t}$ dispersion curve is very steep such that the conduction
time is minimal, except for a short interval of $D$ close to $0$. As a
consequence, we approximate that both $R_{1}$ and $F$ have been propagating
with the minimum conduction time (i.e. maximum speed) $c_{t,\min}$, such that
\begin{align*}
T_{rep,F}(y)  &  \simeq-c_{t,\min}y\\
T_{act,R1}(y)  &  \simeq D_{S1}+c_{t,\min}y\\
D_{R1}(y)  &  =T_{act,R1}(y)-T_{rep,F}(y)\simeq D_{S1}+2c_{t,\min}y
\end{align*}
Substituting these relations in eq. \ref{cond_R2s} yields%
\begin{equation}
T_{rep,R1}\left(  y\right)  \simeq D_{S1}+c_{t,\min}y+A^{SP}\left(
D_{S1}+2c_{t,\min}y\right)  \label{eq_trep_r1}%
\end{equation}
Taking the spatial derivatives of eq. \ref{eq_trep_r1}\ yields%
\[
\frac{dT_{rep,R1}}{dy}=c_{t,\min}\left(  1+2\left.  \frac{dA^{SP}}%
{dD}\right\vert _{D_{R1}\left(  y\right)  }\right)  .
\]
which, thanks to eq. \ref{eq_trep_r1}, gives%

\[
\left.  \frac{dA^{SP}}{dD}\right\vert _{D_{R1}\left(  y_{0}\right)  =D_{\max}%
}=\frac{c_{t,\max}-c_{t,\min}}{2c_{t,\min}}%
\]
The existence and location of the critical point depends on the slope of the
restitution curve, and on the relative difference between the maximum and
minimum conduction time. From\ the $c_{t}(D)$ given by eq. \ref{eq_dispersion}
, $c_{t,\max}\simeq3.13c_{t,\min}$, and%
\begin{equation}
\left.  \frac{dA^{SP}}{dD}\right\vert _{D_{R1}\left(  y_{o}\right)  =D_{\max}%
}\simeq1.065. \label{eq_Dmax}%
\end{equation}
Using eq. \ref{eq_restitution} to solve this equation, we obtain that
$D_{\max}\approx D_{S1}+2c_{t,\min}y_{0}\simeq70$ ms. This result means that,
if $D_{S1}>D_{\max}\simeq70$ ms, $\left\vert dT_{rep,R1}/dx\right\vert
<c_{t,\max}$\ everywhere between $x_{c}$ and $x_{s}$, and $R_{2}$ cannot be
blocked. It means also that if $D_{S1}<D_{\max}$, there is a critical point
$x_{c}<x_{0}<x_{s}$ whose position depends on $D_{S1}$ but is independent of
$L$. The last step is to determine the maximum $\Delta_{S1-S2}$ ($\Delta
_{S1-S2,\max}$) to get a block of $R_{2\text{ }}$ at the $x_{0}$. Assuming
again that $R_{2}$ travels with the minimum conduction time, the first
condition of eq. \ref{cond_R2} becomes%
\[
D_{S1}+c_{t,\min}y+A^{SP}\left(  D_{\max}\right)  =D_{S1}+\Delta_{S1-S2,\max
}+c_{t,\min}y
\]
reducing to the limit%
\[
\Delta_{_{S1-S2}},_{\max}=A^{SP}\left(  D_{\max}\right)
\]
The maximum $\Delta_{S1-S2}$ for the block of $R_{2\text{ }}$is independent of
$L$ and of $D_{S1}$, provided that .$D_{S1}<D_{\max}$. For $S2$ to induce
propagation $\Delta_{S1-S2}$, must also be $\geq APD(D_{S1})$, which is the
refractory period at $x_{s}$. In summary, for all $0\leq D_{S1}\leq D_{\max
}\simeq70$ ms, $R_{2}$ blocks if\ $APD^{SP}(D_{S1})\leq\Delta_{_{S1-S2}}\leq
A^{SP}\left(  D_{\max}\right)  $.

The main approximation used herein is that both $R_{1}$ and $R_{2}$ propagate
everywhere with the minimal propagation time. With regard to $R_{1}$, the
error introduces by the approximation is minimal unless $D_{S1}\simeq0$. For
$R_{2}$, the conduction time is obviously underestimated when its activation
time comes close to $T_{rep,R1}$.\ The approximation thus overestimates
$D_{\max}$, whose value is around 50 ms for the ID and IL \ model, compared to
the $D_{\max}=70$ ms provided by the approximation.

\section{Block of $A_{1}$ after one rotation on the loop\label{appendix_B}}

Appendix \ref{appendix_A} shows that $\forall D_{S1}\in\lbrack0,D_{\max}]$,
$R_{2}$ exist and is blocked if $\Delta_{_{S1-S2}}\in\lbrack APD^{SP}%
(D_{S1}),A^{SP}\left(  D_{\max}\right)  ],$ a set of conditions that do not
depend on $L$. The next event is the annihilation $A_{1}$ when it returns near
$x_{s}$ and hits the refractory tail left $A_{2}$. Using $T_{rep,F}(x_{s})$ as
a reference time $A_{1}$ comes back to $x_{s}$ at the time $D_{S1}+\tau_{A1}$,
in which $\tau_{A_{1}}$ is the time taken by $A_{1}$ to propagates over the
loop on its first turn. It is blocked if the system is still refractory, which
means
\begin{equation}
T_{rep,S2}>D_{S1}+\tau_{A_{1}} \label{cond_block_A1}%
\end{equation}
Neglecting the effect of coupling on repolarization, $T_{rep,S2}$ is
approximated as:
\begin{equation}
T_{rep,S2}\simeq D_{S1}+\Delta_{S1-S2}+A^{SP}(D_{S2}) \label{Trep_S2_blockA1}%
\end{equation}
in which $D_{S2}$, the diastolic interval associated to $S_{2}$, is estimated by%

\begin{equation}
D_{S2}\simeq\Delta_{S1-S2}-A^{SP}(D_{S1}) \label{eq_ds2ds1}%
\end{equation}
Substituting eq. \ref{Trep_S2_blockA1} and \ref{eq_ds2ds1} in eq.
\ref{cond_block_A1} yields%
\begin{equation}
\Delta_{S1-S2}+A^{SP}(\Delta_{S1-S2}-A^{SP}(D_{S1}))>\tau_{A_{1}}\text{,}
\label{cond_block_A1_rr}%
\end{equation}
in which $\tau_{A_{1}}$ is the only non- local term. For most ionic models and
experimental preparations, $\tau_{A1}$ is a monotonic decreasing function of
$D_{S1}$ (as in fig. 4 of ref. \cite{Comtois2002_Re}). The effect of the
prematurity on the return cycle comes from the steepness of the $c_{t}$
dispersion curve (eq. \ref{eq_dispersion}) which is close to $c_{t,\min}$ as
soon as $D$ is greater than a few tenths of ms. Therefore, the prolongation of
the return cycle depends on the limited region beyond $x_{s}$ over which
$A_{1}$ does not propagate at the maximum speed. Hence, we write
\begin{equation}
\tau_{A1}\left(  D_{S1},L\right)  \approx Lc_{t,\min}+f\left(  D_{S1}\right)
\label{eqtau1}%
\end{equation}
where $f$ is maximum for $D_{S1}=0$ and
\begin{align*}
\frac{df}{dD_{S1}}  &  <0,\\
\lim_{D_{S1}\longrightarrow\infty}f  &  =0
\end{align*}
The condition \ref{cond_block_A1_rr} for the block of $A_{1}$ becomes%

\begin{equation}
\Delta_{S1-S2}+A^{SP}(\Delta_{S1-S2}-A^{SP}(D_{S1}))>Lc_{t,\min}+f\left(
D_{S1}\right)  \text{.} \label{lim_A1_bl_final}%
\end{equation}
Consider
\begin{equation}
\Delta_{S1-S2}^{\lim}+A^{SP}(\Delta_{S1-S2}^{\lim}-A^{SP}(D_{S1}))=Lc_{t,\min
}+f\left(  D_{S1}\right)  \label{eq_condf}%
\end{equation}
as the minimum value of $\Delta_{S1-S2}$ to get a block of $A_{1}$.

Lets consider next the case where $D_{S1}<D_{\max}$ and $\Delta_{S1-S2}%
=A^{SP}(D_{S1})$, for which $D_{S2}=0$. The condition for the block of $A_{1}$ becomes%

\[
A^{SP}(D_{S1})+A^{SP}(0)=Lc_{t,\min}+f\left(  D_{S1}\right)
\]
which cannot be fulfilled since both $D_{S1}$ and $A^{SP}(0)$ are smaller than
the diastolic interval of the free reentry. The block of $A_{1}$ will occur
from the minimum $\Delta_{S1-S2}>A^{SP}(D_{S1})$. This explains why the
dynamical regime [$A_{1}^{1}\rightarrow,A_{2}^{1}\dashv$] is found in the
lower portion of the [$D_{S1},\Delta_{S1-S2}$] area in which $R_{2}$ blocked,
as seen in figure \ref{fig_map_super}.

The condition for the block of $R_{2}$ are that $D_{S1}\in\lbrack0,D_{\max}]$
and $\Delta_{_{S1-S2}}\in\lbrack APD^{SP}(D_{S1}),A^{SP}\left(  D_{\max
}\right)  ]$. Since the right hand side of eq. \ref{eq_condf} is a growing
function of $L$, $\Delta_{S1-S2}^{\lim}$ for any fixed value of $D_{S1}$ must
also increase with $L$ until reaching $A^{SP}\left(  D_{\max}\right)  $.
Hence, there is a limit value of $L$ from which $A_{1}$cannot be blocked on
its first return. For $D_{S1}^{\lim}(L)$, the value of $D_{S1}$ for which
$\Delta_{S1-S2}^{\lim}=APD^{SP}(D_{\max})=\Delta_{S1-S2,\max}$, eq.
\ref{eq_condf} becomes%
\[
\Delta_{S1-S2,\max}+A^{SP}(\Delta_{S1-S2,\max}-A^{SP}(D_{S1}^{\lim
}(L))=Lc_{t,\min}+f\left(  D_{S1}^{\lim}(L)\right)
\]

Shortening $D_{S1}^{\lim}(L)$ \ increases both the left and right side of the
equation, such that the value of $D_{S1}^{\lim}(L)$ depends on the balance
between the slope of $df/dD$ (i.e. the change in the return cycle) and
$dA^{SP}/dD$ (the change in repolarisation time at $x_{s}$). The derivative of
the equation with respect to $L$ gives:%
\[
\left(  -\frac{df}{dD_{S1}}-a_{D_{S1}}^{\prime}a_{D_{S2}}^{\prime}\right)
\frac{dD_{S1}^{\lim}}{dL}=c_{t,\min}%
\]
where $a_{D_{0}}^{\prime}=\left.  dA^{SP}/dD\right\vert _{D=D_{0}}$. In the
MBR model, $\left\vert \partial f/\partial D_{S1}\right\vert <1,$ whereas
$dA^{SP}/dD>1$ at low $D$ value, such that $dD_{S1}^{\lim}/dL<0,$ as observed
in the numerical simulations.

\section{Block of $A_{2}$ after one rotation on the loop\label{appendix_C}}

Appendix \ref{appendix_B} \ shows that $A_{1}$ is blocked only over a subset
of the [$D_{S1},\Delta_{_{S1-S2}}$] area for which $R_{2}$ exist and is
blocked (i.e. $D_{S1}\in\lbrack0,D_{\max}]$, $\Delta_{_{S1-S2}}\in\lbrack
APD^{SP}(D_{S1}),A^{SP}\left(  D_{\max}\right)  ]$). With respect to
$\Delta_{_{S1-S2}}$, the lower bound of this subset is greater than
$APD^{SP}(D_{S1})$ and increases with $L$, while the upper bound remains
constant at $A^{SP}\left(  D_{\max}\right)  $. With respect to $D_{s1},$the
variation of the limits of this subset is given by a complex expression that
depends on the slope of both the restitution and dispersion curves. For the
MBR model, the upper $D_{S1}$ of the subset decreases toward 0 as $L$ is
increased. The last step is to obtain the conditions for the block of $A_{2.}$

$A_{2}$ blocks between $x_{c}$ and the locus where $R_{2\text{ }}$has stopped
when it hits the refractory tail left by $A_{1}$. Using the $y$ coordinate,
the condition for the block of $A_{2}$ is that there is a point where%

\begin{equation}
T_{rep,A1}\left(  y\right)  >T_{act,A2}\left(  y\right)  \label{condA2_1}%
\end{equation}
\newline$T_{rep,A1}(y)$ is approximated as $T_{act,A1}\left(  y\right)
+A^{SP}(D_{A1}\left(  y\right)  ,$where $D_{A1}\left(  y\right)
=T_{act,A1}\left(  y\right)  -T_{rep,R1}\left(  y\right)  $ since $R_{2}$ has
not propagated in this region. Using eq. \ref{eq_trep_r1}\ of appendix
\ \ref{appendix_A} \ for $T_{rep,R1}(y)$\ and eq. \ref{eqtau1} of appendix
\ref{appendix_B} for $T_{act,A1}\left(  y\right)  $\ yields%

\begin{equation}
T_{repol,A1}=D_{S1}+Lc_{t,\min}+f\left(  D_{S1}\right)  -c_{t,\min}%
y+A^{SP}(D_{A1}\left(  y\right)  \label{eq_trepola1}%
\end{equation}

\[
D_{A1}\left(  y\right)  =Lc_{t,\min}+f\left(  D_{S1}\right)  -2c_{t,\min
}y-A^{SP}\left(  D_{S1}+2c_{t,\min}y\right)
\]
\newline$T_{act,A2}\left(  y\right)  =D_{S1}+\Delta_{_{S1-S2}}+\tau_{A2}(y)$
where $\tau_{A2}$ is the time needed for $A_{2}$ to travel from the
stimulation site to $y$. For $A_{1}$, we have assumed in \ref{appendix_B} that
the prolongation of the return cycle was occurring mainly in a short region
around $x_{s}$ beyond which $A_{1}$ was travelling at maximal speed. The
situation is different for $A_{2}$. As it could be seen in fig.
\ref{fig_bloc_aa} A, the $D$ associated to $A_{2}$ as it propagates away form
$x_{s},$meaning that it speed of propagation diminishes. Nevertheless, we write%

\begin{equation}
\tau_{A2}(y)=Lc_{t,\min}+g\left(  D_{S1,}D_{s2}\right)  -c_{t,\min}y
\label{eq_tact2}%
\end{equation}
\newline With eq. \ref{eq_trepola1} and \ref{eq_tact2}, the condition
\ref{condA2_1} \ for the block $A_{2}$ after one rotation becomes
\begin{equation}
f\left(  D_{S1}\right)  +A^{SP}(D_{A1}\left(  y\right)  )>\Delta
_{S1-S2}+g\left(  D_{S1},D_{S2}\right)  \label{lim_A2_bl_final1}%
\end{equation}

To get the [$A_{1}^{1}\dashv,A_{2}^{1}\dashv$] block, the conditions given by
eq. \ref{lim_A1_bl_final} and eq. \ref{lim_A2_bl_final1} must both be
fulfilled, leading to the final condition:%

\[
A^{SP}\left(  \Delta_{S1-S2}-A^{SP}\left(  D_{S1}\right)  \right)
>Lc_{t,\min}+g\left(  D_{S1},D_{S2}\right)  -A^{SP}\left(  D_{A1}\left(
y\right)  \right)  .
\]
with the supplementary constraint that $\Delta_{S1-S2}$ and $D_{S1}$ must
remain in the interval where $R_{2}$ is block. Since both $\Delta_{S1-S2}$ and
$D_{S1}$ are bound, there is a limiting $L$ value beyond which [$A_{1}%
^{1}\dashv,A_{2}^{1}\dashv$] block cannot occur. For the IL and ID model,
$\ $we found that $A_{2}$ was always stop when $A_{1}$ was blocked, meaning
that eq. \ref{lim_A2_bl_final1} \ was fulfilled whenever eq.
\ref{lim_A1_bl_final} was satisfied. However, the condition depends on the
restitution and dispersion curve and, through $g\left(  D_{S1},D_{S2}\right)
$, the interaction of $A_{2}$ with the spatial profile of $T_{repol}$ left
by $A_{1}$.

\end{document}